\documentclass[fleqn,usenatbib,useAMS]{mnras}

\usepackage{graphicx}	
\usepackage{amsmath}	
\usepackage{amssymb}	
\usepackage{appendix}
\usepackage{multicol}       
\usepackage{bm}              
\usepackage{pdflscape}   

\usepackage{verbatim}
\usepackage{color}
\usepackage[usenames,dvipsnames]{xcolor}
\definecolor{green}{rgb}{0.0, 0.4, 0.0}
\definecolor{forestgreen(web)}{rgb}{0.13, 0.55, 0.13}
\definecolor{green(web)}{rgb}{0.13, 0.55, 0.13}
\definecolor{green}{rgb}{0.0, 0.4, 0.0}

\title[The evolution of merger fraction of galaxies]
{The evolution of merger fraction of galaxies at $\it{z}$ $<$ 0.6 depending on the star formation mode in the AKARI NEP--Wide field }

\author[Eunbin Kim]{Eunbin Kim$^{1}$%
\thanks{Contact e-mail:\href{mailto:ebkim@si-analytics.ai}{ebkim@si-analytics.ai}}%
\thanks{Present Address: SI-analytics, Yusungdaero 1689 gil 70, Yusung-gu, Daejeon, 34047, Korea},
Ho Seong Hwang$^{1,2,3}$,
Woong-Seob Jeong$^{1}$,
Seong Jin Kim$^{3}$,
\newauthor{
Denis Burgarella$^{4}$,
Tomotsugu Goto$^{3}$,
Tetsuya Hashimoto$^{3,5}$,
Young-Soo Jo$^{1}$,}
\newauthor{
Jong Chul Lee$^{1}$,
Matthew Malkan$^{12}$,
Chris Pearson$^{18,19,20}$,
Hyunjin Shim$^{6}$,}
\newauthor{
Yoshiki Toba$^{7,8,9}$,
Simon C.-C. Ho$^{3}$,
Daryl Joe Santos$^{3}$,
Hiroyuki Ikeda$^{10,11}$,}
\newauthor{
Helen K. Kim$^{12}$,
Takamitsu Miyaji$^{13,14}$,
Hideo Matsuhara$^{15,16}$,
Nagisa Oi$^{17}$,}
\newauthor{
Toshinobu Takagi$^{21}$,
Ting-Wen Wang$^{3}$}
\\
$^{1}$Korea Astronomy and Space Science Institute, 776 Daedeokdae-ro, Yuseong-gu, Daejeon 34055, Korea\\
$^{2}$Astronomy Program, Department of Physics and Astronomy, Seoul National University, 1 Gwanak-ro, Gwanak-gu, Seoul 08826, Korea \\
$^{3}$SNU Astronomy Research Center, Seoul National University, 1
Gwanak-ro, Gwanak-gu, Seoul 08826, Republic of Korea\\
$^{4}$Institute of Astronomy, National Tsing Hua University No. 101, Section 2, Kuang-Fu Road, Hsinchu 30013, Taiwan\\ 
$^{5}$Aix Marseille Universit\'e, CNRS, Laboratoire d'Astrophysique de Marselle UMR 7326, 13388 Marseille, France\\
$^{6}$Centre for Informatics and Computation in Astronomy (CICA), National Tsing Hua University, 
101, Section 2. Kuang-Fu Road,\\ Hsinchu, 30013, Taiwan\\
$^{7}$Department of Earth Science Education, Kyungpook National University, 80 Daehakro, Bukgu, Daegu 41566, Korea\\
$^{8}$Department of Astronomy, Kyoto University, Kitashirakawa-Oiwake-cho, Sakyo-ku, Kyoto 606-8502, Japan\\
$^{9}$Academia Sinica Institute of Astronomy and Astrophysics, 11F of Astronomy-Mathematics Building, AS/NTU, No.1, Section 4,\\ Roosevelt Road, Taipei 10617, Taiwan\\
$^{10}$Research Center for Space and Cosmic Evolution, Ehime University, 2-5 Bunkyo-cho, Matsuyama, Ehime 790-8577, Japan\\
$^{11}$National Astronomical Observatory of Japan, 2-21-1 Osawa, Mitaka, Tokyo 181-8588, Japan\\
$^{12}$National Institute of Technology, Wakayama College, Gobo, Wakayama 644-0023, Japan\\
$^{13}$Department of Physics and Astronomy, UCLA, 475 Portola Plaza, Los Angeles, CA 90095-1547, USA\\
$^{14}$Instituto de Astronom\'a sede. Ensenada, Universidad Nacional Aut\'onoma de M\'exico (UNAM), Km 107, Carret. Tij.-Ens.,\\
Ensenada, 22060, BC, M\'exico\\
$^{15}$Leibniz Institut f\"ur Astrophysik Potsdam, An der Sternwarte 16, 14482 Potsdam, Germany\\
$^{16}$Department of Space and Astronautical Science,The Graduate University for Advanced Studies, SOKENDAI, 3-1-1, Yoshinodai,\\
Chuo-ku, Sagamihara, Kanagawa 252-5210, Japan\\
$^{17}$Institute of Space and Astronautical Science, Japan Aerospace Exploration Agency, 3-1-1 Yoshinodai, Chuo-ku, Sagamihara,\\
Kanagawa, 252-5210, Japan\\
$^{18}$Tokyo University of Science, 1-3, Kagurazaka Shinjuku-ku Tokyo 162-8601 Japan\\
$^{19}$RAL Space, STFC Rutherford Appleton Laboratory, Didcot, Oxfordshire, OX11 0QX, UK\\
$^{20}$Oxford Astrophysics, University of Oxford, Keble Rd, Oxford OX1 3RH, UK\\
$^{21}$The Open University, Milton Keynes, MK7 6AA, UK\\
$^{22}$Japan Space Forum, 3-2-1, Kandasurugadai, Chiyoda-ku, Tokyo 101-0062, Japan\\
}

\date{Accepted XXX. Received YYY; in original form ZZZ}
\pubyear{2021}

\begin{document}
\label{firstpage}
\pagerange{\pageref{firstpage}--\pageref{lastpage}}
\maketitle

\clearpage
\begin{abstract}
We study the galaxy merger fraction and its
 dependence on star formation mode in the 5.4 square degrees of the North Ecliptic Pole-Wide field.
We select 6352 galaxies with AKARI \( 9\ \mu \)m detections,
  and identify mergers among them using the Gini coefficient and M$_{20}$ derived from the Subaru/HSC optical images. 
We obtain the total infrared luminosity and star formation rate of galaxies using
  the spectral energy distribution templates based on one band, AKARI  \( 9\ \mu \)m. 
We classify galaxies into three different star formation modes (i.e. starbursts, main sequence, and quiescent galaxies)
  and calculate the merger fractions for each.
We find that the merger fractions of galaxies increase with redshift at $\it{z}$ $<$ 0.6. 
The merger fractions of starbursts are higher than those of main sequence 
  and quiescent galaxies in all redshift bins.
We also examine the merger fractions of far-infrared detected galaxies
  which have at least one detection from $\it{Herschel}$/SPIRE.  
We find that $\it{Herschel}$ detected galaxies have higher merger fraction 
   compared to non-$\it{Herschel}$ detected galaxies, and 
   both $\it{Herschel}$ detected and non-$\it{Herschel}$ detected galaxies
   show clearly different merger fractions depending on the star formation modes.
\end{abstract}



\section{INTRODUCTION}

Galaxy interactions and mergers are thought to play an important role in galaxy evolution, impacting their 
morphologies, gas kinematics, and star formation rates \citep{toom72,  sand88, barn96, cons06}.
Cold dark matter models predict that 
  galaxies have accreted their mass through hierarchical mergers \citep{delu06}.
Mergers are also suspected to trigger luminosity increases, from
  active galactic nuclei (AGNs, \citealt{sand96}) and
  starbursts \citep{barn96,miho96,cox06}.
Enhanced star formation can 
 result from the tidal interactions of the galaxies that
   compress/shock the gas,
 causing it to collapse and form stars \citep{barn04,kim09,sait09}.
These merger-induced starbursts are sometimes observed as 
 luminous/ultra-luminous infrared galaxies (LIRGs/ ULIRGs),
 which have extreme far-infrared (FIR) luminosities of 
10$^{11}$ L$_{\odot}$ and 10$^{12}$ L$_{\odot}$, respectively \citep{sand96}.
Several studies have shown that the infrared luminosity of galaxies is statistically correlated 
with mergers \citep{elli13,lars16}.
These can be seen as morphological disturbances 
  in LIRGs and ULIRGs \citep{elbaz03,hwang07,hwang09,kart10,elli13}.

However, the merger contributions to starbursts in general is still unclear.
At high redshifts, average star formation rates
are higher. 
LIRGs and ULIRGs are often found on the ``main sequence", that is, obeying the correlation between SFR
and stellar mass of typical galaxies at a given redshift \citep{dadd07,elbaz07}.
Thus, we will define starbursts by comparison
with the star formation of other galaxies of a given
stellar mass
in the same redshift bin. 
Previous studies have defined starbursts as galaxies experiencing star formation
  three or four times above the median of the SFRs of
  main sequence of star-forming galaxies \citep{elbaz11,rodi11,schr15}.
In a similar way, we divide galaxies into three different star formation ``modes"
  (i.e. starbursts, main sequence and quiescent galaxies) in each redshift bin.

It is challenging
  to identify a large number of merger galaxies out to large redshifts.  
There are two main methods to identify mergers -- selecting close pairs \citep{bund09,dera09,man16,dunc19}
  or using morphological disturbances \citep{cons03,lotz08,lotz11}.
For example, \cite{kado18} used the Subaru/HSC images to select the merger galaxies at 0.05 $< z <$ 0.45 using visually identified tidal features (e.g. shell or stream features)
However, for using close pairs, spectroscopic velocities for both pairs are needed.
Because spectroscopic observations are expensive,
 merger studies based on them will suffer from incompleteness.
Deep and high-resolution imaging (e.g. {\it Hubble Space Telescope})
  could avoid that incompleteness, but
  the merger fraction from galaxy imaging
  may be ambiguous.
Morphological disturbances presented the late-stage of mergers
  can be determined by visual inspection or 
  discriminate quantitative outliers of morphological disturbances.
Visual inspection is subjective and time-consuming.
High redshift galaxies can also be easily misclassified
  because of wavelength-dependent morphology and 
  surface brightness effects \citep{bohl91,kuch00,wind02,kamp07}.

In general, morphological types of galaxies are classified by the light profiles of galaxies.
The measured profile is the average intensity of a galaxy as a function of radius,
  and can be mathematically fitted (e.g. S{\'e}rsic profile, \citealt{sers63}).
These parametrizations historically have been used to classify galaxies 
  into Ellipticals, Spirals and Irregular galaxies.
There are also non-parametric measures for galaxy classification,
   such as  concentration, asymmetry, clumpiness (CAS, \citealt{cons00,cons03,mena06}).
In addition, there have been many studies of morphological classification using the parameters of Gini coefficient and M$_{20}$ \citep{lotz04,lotz08}. The Gini coefficient is a measure whether the flux of a galaxy is concentrated or spread out (to be formally defined in Section 3), and M$_{20}$ is the the second-order moment of the brightest 20 percent of a galaxy \citep{lotz04,lotz08}. The Gini coefficient is originally used in economics to statistically describe the distribution of wealth within a society. This coefficient was applied to astronomical images to quantify the spread of galaxy light \citep{abra03,lotz08}, which is now widely used for morphological analysis in astronomy.
Between these two approaches (parametric vs. non-parametric), non-parametric measurements may be less impacted by redshift.
Concentration, asymmetry, clumpiness
  are used for merger-finding, 
  but asymmetry measurement is less sensitive to the late-stage of mergers \citep{lotz08}.
The Gini coefficient and M$_{20}$
  are used for wide area surveys \citep{lotz04,lotz10a,lotz10b}.
These parameters are more effective in classifying galaxies and 
  identifying late-stage mergers than concentration and asymmetry,
  and also more robust for galaxies with low signal-to-noise ratios \citep{lotz04,lotz08}.
Therefore, we use the Gini coefficient and M$_{20}$ for identifying mergers,
 for quantitative comparison with other studies, and
  to secure a large sample from our data.
  
\begin{figure}
\centering
\includegraphics[width=0.45\textwidth]{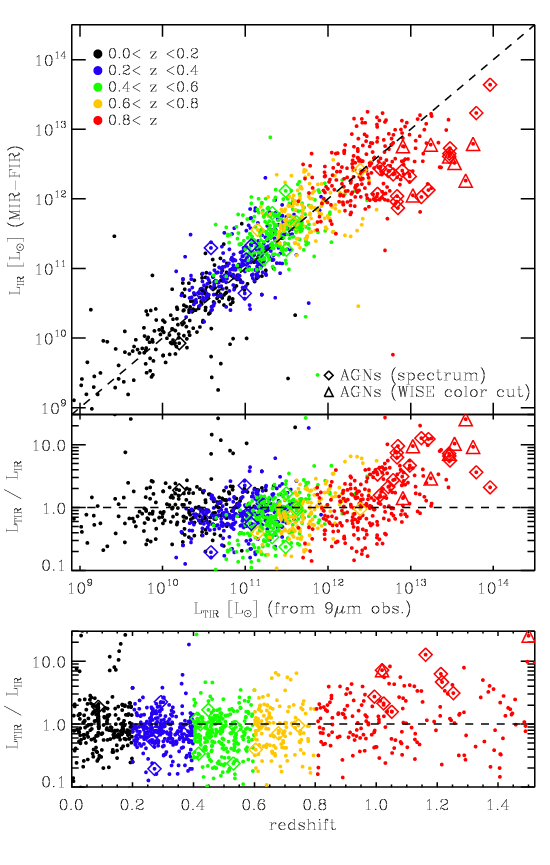}
\caption{L$_{IR}$(MIR-FIR) versus L$_{TIR}$ from AKARI \( 9\ \mu \)m (Top). 
L$_{IR}$(MIR-FIR)  is fitted by CIGALE including at least one FIR band detection \citep{wan20}.
Diamonds and triangles represent AGNs classified with spectra and WISE colour cut, respectively.
(Middle) The ratio of the L$_{TIR}$ to L$_{IR}$(MIR-FIR) as a function of L$_{TIR}$.
(Bottom) The ratio of the L$_{TIR}$ to L$_{IR}$(MIR-FIR) as a function of redshift.
Colour codes represent galaxies in different redshift ranges.
}
\label{fig-lirtot}
\end{figure}

In this paper, we examine the evolution of merger fractions of galaxies with 
  redshifts up to $\it{z}$ = 0.6 in the AKARI North Ecliptic Pole (NEP)--Wide field,
  and the variations in merger fractions of galaxies in three different star-formation modes 
  (i.e. starbursts, main sequence, and quiescent galaxies).
We also study how the FIR detection affects the merger fractions.
In Section 2, we summarise our observational data and sample selection.
Section 3 describes the morphological analysis
to classify the galaxies using the Gini and M$_{20}$.
We present the results and discussion in Sections 4 and 5, respectively.

\section{DATA}
\subsection{Optical images}\label{sec:opt}
We use deep optical images taken with the Hyper Supreme-Cam (HSC) on Subaru 8-m telescope 
  in the AKARI NEP--Wide field covering 5.4 deg$^2$ (\citealt{goto17}, Oi et al. in accepted).
The HSC has a 1.5 deg field of view (FoV) covered with 104 red-sensitive CCDs,
  and the pixel scale is 0.17 arcsec.
It is the largest FoV among the 8-m telescopes,
  and the size of FoV covered the AKARI NEP--Wide field with only four pointings 
  (see Figure 2 in \citealt{goto17}, Figure 1 in Oi et al. in accepted).
The observations of the NEP--Wide field were performed in June 30th 2014 and August 7-10th in 2015. The 5$\sigma$ detection limits are 28.6, 27.3, 26.7, 26.0, 25.6 AB mag, 
  and the median seeings are 0.68, 1.26, 0.84, 0.76, 0.74 for  $\it{g, r, i, z}$ and $\it{y}$-band, respectively.
The total number of identified sources in 5 bands is 3.5 million and more detailed information on the data set is described in Oi et al. (accepted).
 
\subsection{Multi-wavelength Data}

We have used the multi-wavelength data set based on the catalogue of AKARI mid-IR (MIR)
  galaxies newly identified by an optical survey by Subaru/HSC \citep{oi21}.  
The infrared galaxies detected by AKARI's NEP--Wide survey \citep{kim12} 
  were cross-matched against deep HSC optical data, 
  thereafter all available supplementary data over the NEP--Wide field 
  were merged together \citep{kim21}.

Data merging of these two catalogues were carried out by positional matching 
  with the matching radii defined by 3-sigma positional offsets, 
  which are more rigorous than using PSF sizes \citep{kim21}.
This band-merged catalogue has 91,000 objects including
  $\sim$ 70,000 objects detected in N2, N3, N4 bands,
  $\sim$ 20,000 objects detected in S7, S9, L11, L18, L24 bands,
  and is the reference catalogue for our sample selection in this study.
Optical to submillimeter (submm) photometry for AKARI sources are also added.
Original AKARI NEP-Wide field catalogue \citep{kim12} includes
  CFHT/MegaCam ${\it u^*, g', r', i', z'}$ \citep{hwan07},
  Maidanak observatory/ SNUCAM B, R, I-band data \citep{jeon10}
  and KPNO /FLAMINGOS J and H band data \citep{jeon14}.
Supplementally, the observed data from
  CFHT/MegaPrime $\it{u}$-band \citep{huan20},
  CFHT/MegaCam ${\it u^*, g', r', i', z'}$ \citep{oi14,goto18},
  WIRCam Y, J and Ks band \citep{oi14} are
  added to the main catalogue.
The main catalogue is also cross-matched with
  the $\it{WISE}$ catalogue \citep{jarr11},
  $\it{Spitzer}$/IRAC \citep{nayy18} and 
  $\it{Herschel}$/PACS and SPIRE \citep{pear17,pear19}.
This band-merged catalogue adopted spectroscopic redshifts for objects from several  observations,
  which include Keck/DEIMOS \citep{shog18,kim18} and
  MMT/Hectospec and  WIYN /Hydra \citep{shim13}.
Subaru/FMOS \citep{oi17}, GTC (Miyaji et al. in preparation) and 
  the SPICY survey \citep{ohya18} are also included.  
Photometric redshifts are determined \citep{ho21}
   using 26 bands from optical to NIR 
   with the public code LePhare \citep{arno99,ilbert06},
   and the photo-z accuracy is $\sigma$ = 0.053.

\begin{figure}
  \begin{center}
    \includegraphics[width=0.45\textwidth]{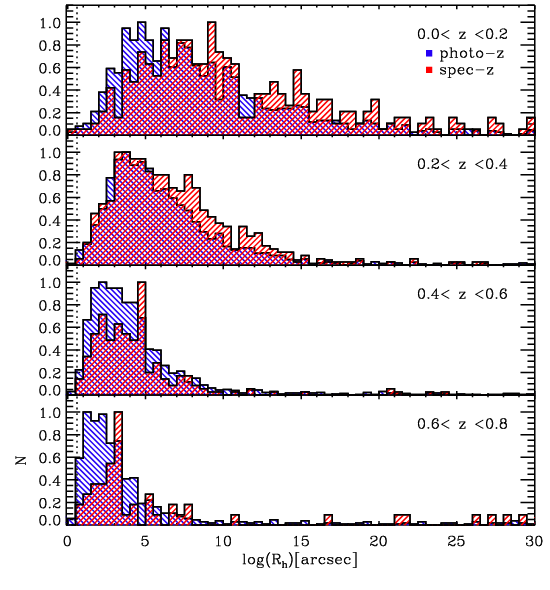}
   \end{center}
\caption{Distribution of radius for galaxies. 
Dotted line represents the seeing size, 0.6 arcsec in the HSC $\it{i}$-band image. 
Blue and red histograms represent the galaxies with photo-z and spec-z. }
\label{fig-rhalf}
\end{figure}
\begin{figure}
  \begin{center} 
    \includegraphics[width=0.45\textwidth]{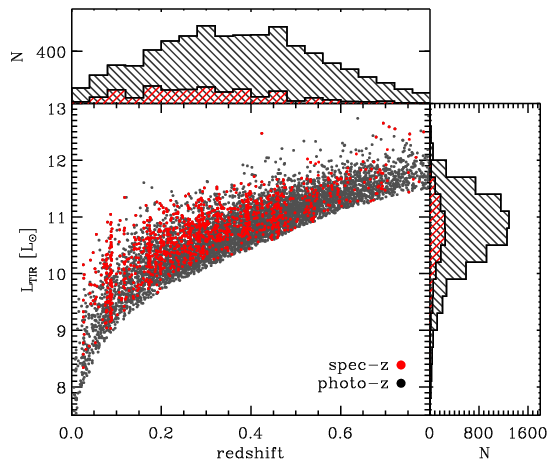}
   \end{center}
\caption{Total infrared luminosity distribution as a function of redshift for 9-$\micron$ selected galaxies. 
Black circles and red crosses represent the galaxies with spec-z and photo-z.  }
\label{fig-z}
\end{figure}
  
\subsection{Physical Parameters of Galaxies}

We derive the total infrared luminosity ($L_{\rm TIR}$, 8--1000 \micron)
  using a set of template spectral energy distributions (SEDs) of main sequence galaxies in \citet{elbaz11}
  with each of AKARI bands, S7, S9, S11, L15, L18 and L24.
They defined a typical IR SED for main sequence
  galaxies using $\it{Herschel}$ data,
  this SED could extrapolate the total IR luminosity
  for galaxies that only one measurement exists.
Although $L_{\rm TIR}$ derived with FIR data could be more accurate than that without FIR data, the latter case enables us to secure a large number of samples \citep{calz10,gala13}.

To assure the validity of the $L_{\rm TIR}$ from one-band, 
   we compare the $L_{\rm TIR}$ with infrared luminosity  ($L_{\rm IR}$
   derived from MIR-FIR bands, \citealt{wan20}). 
They calculated the $L_{\rm IR}$ using SED modeling code CIGALE \citep{burg05,noll09}
  with 36 bands ranging from optical to submm bands, which is represented as a sum of dust and AGN activities.
 
As seen in Figure \ref{fig-lirtot}, we compared the difference between $L_{\rm TIR}$ from one-band and $L_{\rm IR}$
  from MIR--FIR bands as a functions of $L_{\rm TIR}$ and redshift.
$L_{\rm TIR}$ and $L_{\rm IR}$ up to around $L_{\rm TIR}$ = 10$^{13}$ L$_{\odot}$ show
  good agreement below $\it{z}$ $\sim$ 0.8, 
  which may originate from the effect of galaxy evolution on the SED over the cosmic time.
The relation between the total IR luminosity and one-band IR lunminosity
  could have discrepancy as redshift inceases
  depending on which IR-band is used (e.g. 24 vs 8 \micron, \citealt{elbaz11}).
Since such a trend is commonly found in every AKARI MIR band (7 -- 24 \micron),
  we used 9 $\micron$ band (S9) for sample selection 
  because of the largest number of sample.
In the middle panel, the standard deviation in $L_{\rm IR}$/$L_{\rm TIR}$ of galaxies except AGNs is 0.71.  
A detailed sample selection is described in Section 2.4 and Table \ref{tab:sam}.

Since the $L_{\rm TIR}$ is usually used as good star formation indicator, we calculate the $L_{\rm TIR}$ from the initial mass function (IMF).  
We adopt the \citet{salp55} initial mass function (IMF), which is also used for calculating $L_{\rm IR}$ with CIGALE in \citet{wan20}. 
The star formation rates are calculated from the formula (12) in \citet{kenn12} which is
\begin{equation}
\rm{log \dot{ M_{*}} (M_{\odot} \rm{year^{-1}}) = {log} L_x - {log} C_x} ,
\end{equation}

where L$_x$ units are [ergs s$^{-1}$] and logC${_x}$ = 43.41 
adopting the calibration factors from \citet{hao11,murp11}. 
The stellar mass ($M_*$) is derived from SED fitting with LePhare \citep{arno99,ilbert06} using 13 multi-wavelength data of
  CFHT/MegaCam $\it{u'}$, Subaru/HSC $\it{g, r, i, z, Y}$, 
  CFHT/Wircam $\it{J, Ks}$,
  and AKARI N2, N3, N4, S7, S9 bands.
We convert $M_*$ from LePhare based on \citet{chab03} IMF
  to $M_*$ based on \citet{salp55} IMF by dividing by a factor of 0.63
  to fairly compare with others (e.g. \citealt{schr15,pear18}).

The star forming galaxies in the main sequence
follow an empirical power-law relation between the SFR and stellar mass. 
However, \citet{smer18,drai20} showed that quiescent
galaxies have considerable scatter on this relation 
when they compared different indicators such as L$_{TIR}$, H$\alpha$, neon lines, etc. 
Thus, considering to adopt separate
conversion factor for deriving the SFR for each different star formation mode 
might improve the relation. 
However, because we have constrained galaxies with the range of 9.0 $<$ log(M$_*$/M$_{\odot}$) $<$11.5 including relatively massive quiescent galaxies, 
we do not apply the separate conversion factor for quiescent galaxies in this paper.

The star forming galaxies show tight correlations
  between L$_{TIR}$ and SFR (e.g., \citealt{kenn12,hwang10,hwang12}). 
However, this tight correlation can break down for quiescent galaxies, 
  especially for those with low infrared luminosities (e.g., \citealt{smer18,drai20}. 
Thus, it is conceivable that this difference might affect our results.
However, the infrared luminosities of our sample galaxies (even for quiescent galaxies) 
  are generally high enough, the impact of this different correlation is insignificant.

\begin{table}
\centering \caption{Number of Galaxies with 5-sigma detection in different AKARI/IRC bands}
\begin{tabular}{c|cccccc}
\hline
Band & S7 & S9  & S11 & L15 & L18 & L24  \\

\hline
Total & 5007& 9076 & 9099 & 8592 & 10133 & 2384 \\
spec-z & 1022 & 1417 & 1388 & 1117 & 1186 & 532 \\
photo-z & 5003 & 9072 & 9096 & 8589 & 10130 & 2382 \\
\hline
Total  (z$<$0.8) & 3702 & 7236 & 7377 & 4640 & 5068 & 1349 \\
spec-z & 861 & 1239 & 1220 & 927 & 971 & 443 \\
photo-z & 3659 & 7173 & 7317 & 4580 & 5010 & 1320 \\
\hline
Total (z$<$0.6) & 3407 & 6425 & 6200 & 3392 & 3805 & 1150 \\
spec-z & 820 & 1169 & 1137 & 849 & 893 & 413 \\
photo-z & 3348 & 6331 & 6107 & 3307 & 3718 & 1108 \\
Herschel detection & 739 & 1048 & 1051 & 805 & 853 & 467  \\
\hline
\end{tabular}
\label{tab:sam}
\end{table}
\begin{figure*}
\centering
\includegraphics[width=0.7\textwidth]{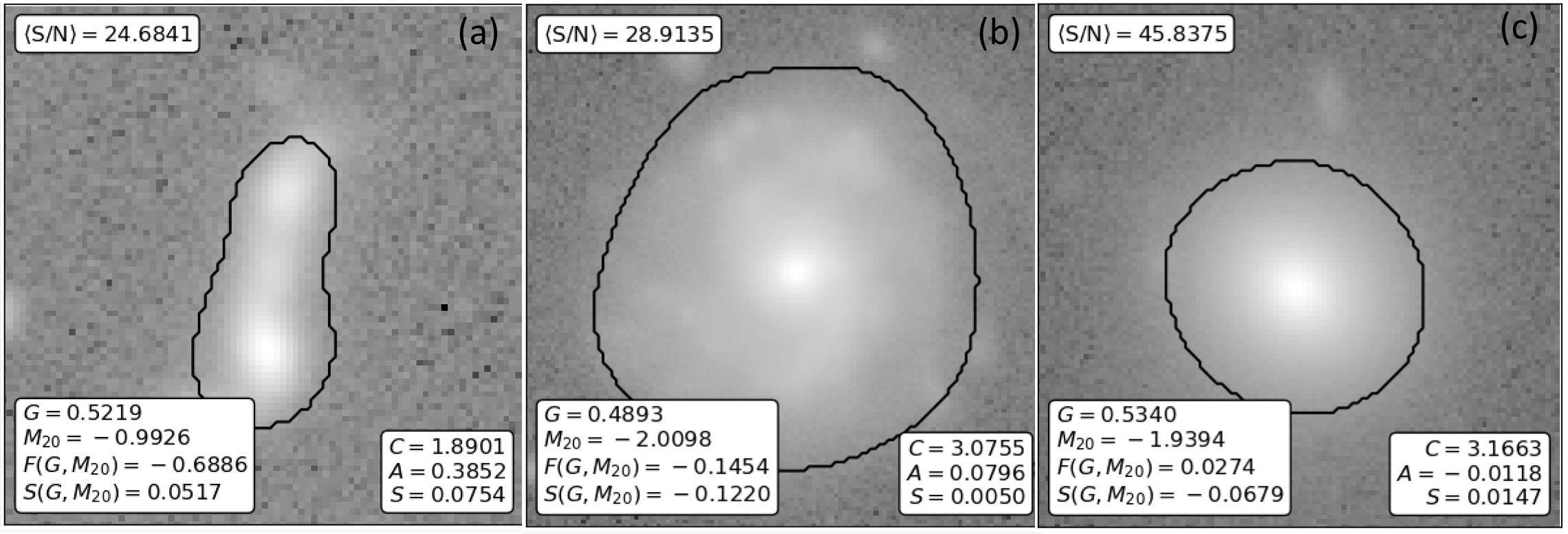}
\caption{Example of the morphological measurements for the galaxy performed by $\texttt{statmorph}$.
(a), (b), (c) are example images of mergers, Spiral, and Elliptical galaxies, respectively.
Black solid contours represent segmentation maps and
text labels represent the values of the Gini, M$_{20}$ and CAS.
F(G,M$_{20}$) and S(G,M$_{20}$) show that bulge statistic and merger statistic.
Detailed parameter description is in \citet{rodr19}.
}
\label{fig-statm}
\end{figure*}

\begin{figure}
\centering
\includegraphics[width=0.48\textwidth]{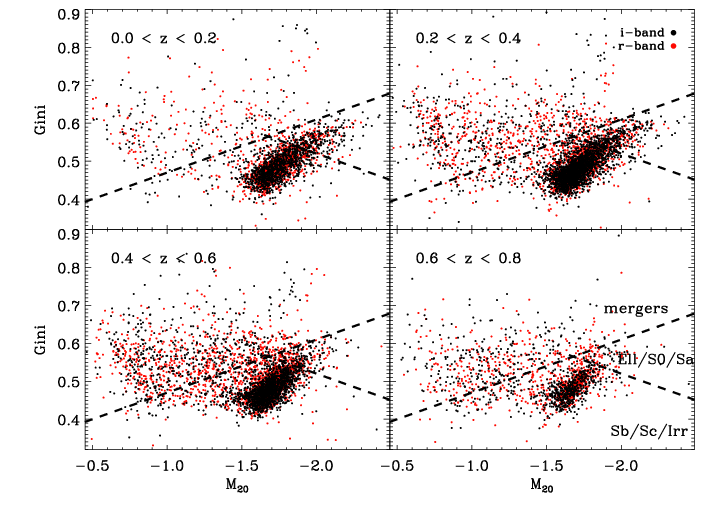}
\caption{The Gini--M$_{20}$ diagram for galaxies at different redshift ranges.
Red and black dots represent galaxies in $\it{r}$ and $\it{i}$ band, respectively.
Dashed lines separate the regimes according to their morphological types; mergers in above the dashed line, Ellipticals in right regime, and Spirals in below the dashed line.
}
\label{fig-class}
\end{figure}

\subsection{Sample Selection} \label{sample}

Considering the total number of detected sources
  at each AKARI band in the band-merged catalogue,
  we selected the S9 (9 $\micron$) band for our study (see Table \ref{tab:sam}).
The total number of galaxies in Table \ref{tab:sam} presents
 the number of galaxies with redshift information 
 estimated from either spectroscopic or photometric measurement.
Although the total number of galaxies with L18 is the largest in the whole redshift range,
  we select 9 $\micron$ detected galaxies for our sample,
  because the number of galaxies is the largest at the redshift below $\it{z}$ $=$ 0.6
  where we finally analyse the data.
Note that the total number of 9 $\micron$ detected galaxies is 9,076 and
  it is reduced to be 7,236 and 6,425 at $\it{z}$ $<$ 0.8 and  $\it{z}$ $<$ 0.6, respectively.
To examine the contribution from AGNs in our analysis,
  we overplot 190 AGNs in Figure \ref{fig-lirtot}, 
  which were identified by Baldwin-Phillips-Terlevich (BPT) emission-line ratio diagrams \citep{shim13}.
In addition, 30 IR-bright AGNs are found through WISE W1 - W2 and W2 - W3 colour-colour diagrams
  with the criteria of \cite{jarr11} and \cite{mate12}.
Figure \ref{fig-lirtot} shows that AGNs are significantly off 
  the linear correlation of $L_{\rm TIR}$ and $L_{\rm IR}$.
This is because AGN-dominant galaxies have higher MIR luminosities 
  compared to star-forming galaxies \citep{spin95}.
Due to these templates are based on star-forming galaxies,
  we remove these 219 AGNs (73 AGNs in $\it{z}$ $<$ 0.6) from the further analysis,
  and end up having 6,352 galaxies at $\it{z}$ $<$ 0.6. 

Figure \ref{fig-rhalf} shows the normalised histogram of distributions of 
  seeing corrected half light radius ($R_h$) for galaxies 
  in each redshift bin, from top to the bottom.
We separate the redshift into the range of 0.2 bin and
  find most of galaxies in size are larger than the seeing 0.6 arcsec 
  in the HSC $\it{i}$-band in all redshift bins.
Figure \ref{fig-z} shows the distribution of sample galaxies on the redshift and the total infrared luminosity. 
Red and black circles represent the sample with spectroscopic and photometric redshifts, respectively.
Upper and right panels show histograms of redshift and $L_{\rm TIR}$, respectively.

\section{Measurement of morphological parameters}

The morphological parameters allow us to classify the galaxy types. In order to quantify galaxy morphologies, 
we used the Gini coefficient and M$_{20}$ classification method \citep{lotz04}. 

The Gini coefficient is a statistical measure of distribution of income
  in a population in economics, and recently has applied to astronomy as well \citep{abra03,lotz04}.
The Gini can be computed sorting the $\it{f_i}$ pixel value increasing order as 
\begin{equation}
 \rm{G= \frac{1}{|\overline{f}|n(n-1) }\sum_{i=1}^{n} (2i -n -1)|f_i|}
\end{equation} 
where $\it{\overline{f}}$ is the mean over the pixel values and 
 $\it{n}$ is the number of pixels.
If all the flux of a galaxy is concentrated in one pixel, G = 1,
  while a galaxy has a homogeneous surface brightness, G = 0 \citep{glas62}.

The M$_{20}$ is the second order moment of brightest regions of a galaxy. 
The brightest 20 $\%$ of the light is normalised 
  to the total second-order central moment, M$_{tot}$ \citep{lotz04}.
These are defined as 
\begin{equation}
\rm{M_{tot} = {\sum_{i}^{n} M_i}  = \sum_{i}^{n} f_i[(x_i - x_c)^2 + ( y_i - y_c)^2]},
\end{equation}

\begin{equation}
\rm{M_{20} = \rm{log}_{10}{\frac {\sum_{i} M_i}{M_{tot}}, while \sum_{i} f_i <0.2 f_{tot}}},
\end{equation}
where $\it{f_i}$ is the pixel flux value and $\it{x_c, y_c}$ is the galaxy centre.
The centre is the point where $\it{M_{tot}}$ is minimised.
The M$_{20}$ is anti-correlated with concentration; 
  low M$_{20}$ represents highly concentrated galaxy.

We derive the non-parametric Gini and M$_{20}$  
  using $\texttt{statmorph}$ python code \citep{rodr19}
  on galaxies in cutouts of $\it{r}$ and $\it{i}$-band images.
It constructs a segmentation map for Gini measurements 
  to be insensitive to dimming surface brightness for distant galaxies \citep{lotz04}.
The image of a galaxy is convolved with the Gaussian kernel $\sigma$ = r$_{petro}$/5,
  where r$_{petro}$ is the Petrosian radius.
The mean surface brightness within the r$_{petro}$ is used to define threshold of flux,
  then the pixel value above the threshold is assigned to the galaxy in the segmentation map.
Both the Gini and M$_{20}$ are calculated on the segmentation map. 
Figure \ref{fig-statm} shows the examples of segmentation maps of 
  three galaxies with measured Gini and M$_{20}$.
It should be noted that the high-column density of dust 
  could impact the morphological classification of galaxies in the Gini-M$_{20}$ space \citep{lotz08}. 
To briefly test this effect, we examine the distributions of Gini and M$_{20}$
  for Herschel detected and non-detected galaxies. 
Because the Herschel detection requires larger submm flux densities 
  (i.e. larger amount of dust than those with similar M$_*$/SFRs/T$_{\rm{dust}}$; \citealt{hild83}, 
  this comparison can show the impact of dust on the morphological measurements.
The comparison does not show any systematic differences of Gini and M$_{20}$ estimates 
  between the two samples (not shown here),
  which is supported by the Kolmogorov-Smirnov test with high significance levels (p $<$ 0.35).  
We therefore do not think that the dust introduces a systematic bias 
  in our measurements of Gini and M$_{20}$ parameter.

As \cite{lotz08} proposed criteria to separate galaxies into three galaxy types
  on the  Gini - M$_{20}$ diagram using galaxies at 0.2 $<$ $\it{z}$ $<$ 1.2,
  we adopt the classification criteria from the equation (4) of \citet{lotz08}
  to divide galaxies into mergers, Spirals and Ellipticals 
  on the Gini and $M_{20}$ diagram: 
  Mergers: G $>$ --0.14 M$_{20}$ + 0.33, E/S0/Sa: G $<$ --0.14 M$_{20}$ + 0.33, and G $>$ 0.14 M$_{20}$ + 0.80,
  Sb/Sc/Irr: G $\leq$ --0.14 M$_{20}$ + 0.33, and  G $\leq$ 0.14 M$_{20}$ + 0.80.
  
Figure \ref{fig-class} shows the G--M$_{20}$ distribution of 
  our sample on $\it{r}$-, $\it{i}$-band images for different redshift bins.
We find that the distribution of these two morphological parameters
  for galaxies on $\it{r}$-, $\it{i}$-band images are not significantly different.
We derive morphological parameters for both $\it{r}$- and $\it{i}$-band images
  to select the one that gives similar rest-frame wavelengths 
  for the comparison of galaxies at different redshifts.
Therefore, we adopt the parameters from $\it{r}$-band images for the galaxies at $\it{z}$ $<$ 0.2, 
  and from $\it{i}$-band images for those at $\it{z}$ $>$ 0.2. 
To verify our morphological classification based on the measurements of Gini and M$_{20}$, we also conduct the visual inspection of the optical images of all the galaxies in our sample. We find that only 1$\%$ of the galaxies classified as mergers in our sample turn out to be spirals, and 1.7$\%$ of ellipticals and 0.2 $\%$ of spirals based only on the G-M$_{20}$ classification are mergers. This contamination is small enough to have no significant impact on our result. We therefore decided to keep the results based on the automated classification based on the estimates of Gini and M$_{20}$ to avoid any possible subjective misclassification based on visual classification, especially for faint galaxies.
Also, since the FWHM of a point source in the i-band images is 0.84 arcsec corresponding to $\sim$ 3.8 kpc at our median redshift (i.e. z $\sim$ 0.3),
we could hardly find patchy features of star formation in the galaxy images that could affect the Gini during visual inspection of galaxies.

\begin{figure}
\centering
\includegraphics[width=0.45\textwidth]{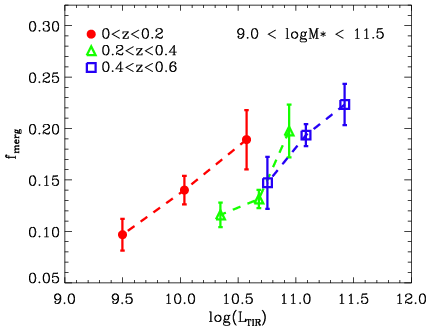}
\caption{Merger fractions as a function of L$_{TIR}$ at different redshift. Filled red circle, open green triangle, open blue rectangle represent galaxies at 0.0 $< \it{z} <$ 0.2, 0.2$< \it{z} <$ 0.4, 0.4$< \it{z} <$ 0.6, respectively.}
\label{fig-add}
\end{figure}

\begin{figure*}
\centering
\includegraphics[width=0.6\textwidth]{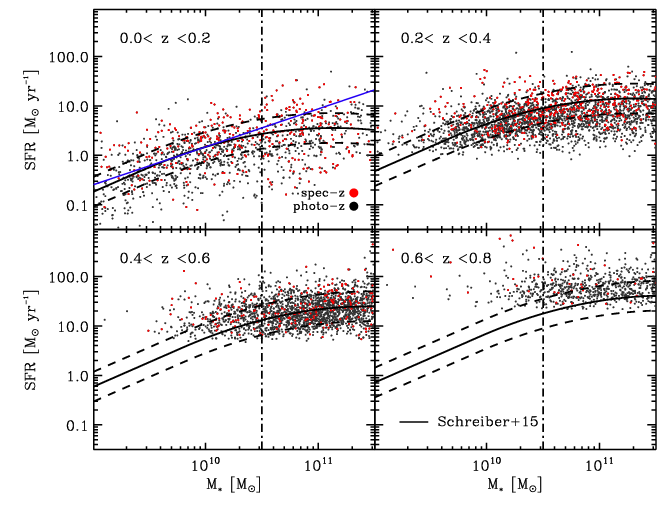}
\caption{ Star formation rates for 9-$\micron$ selected galaxies as a function of stellar mass at different redshift range.
Solid line represents the average SFR for main sequence galaxies \citep{schr15} and 
upper and lower dashed lines represent a factor of two above and below of this fit at each redshift range.
Dash-dotted line represents log(M$_*$) = 10.5. 
Blue solid line shows the best fits for the main sequence galaxies in SDSS \citep{elbaz07}.
Red and black dots represent the galaxies with spec-z and photo-z, respectively.}
\label{fig-sfmode}
\end{figure*}

\section{Results}

It has been well known that the infrared luminosity of galaxies is closely related to the merger activity of galaxies \citep{hwang07, elli13,lars16}. However, the situation can differ if we consider a wide range of redshift. For example,  Figure \ref{fig-add} shows the merger fraction of galaxies in our sample as a function of infrared luminosity at different redshift ranges. 
As expected, the merger fraction increases with L$_{TIR}$ for a given redshift range. However, because the merger fraction could be different depending on the redshift despite similar L$_{TIR}$, we examine the merger fraction focusing on star formation mode.

\subsection{Merger fractions of galaxies at different star formation modes}
The relation between star formation rate and stellar mass of galaxies is tightly 
related to the star formation mode. To investigate the cosmic evolution out to $\it{z}$ $\sim$ 1 
 over the star formation mode, we divide our sample at each redshift bin (see Figure \ref{fig-sfmode}).
We adopt the average SFR of main sequence galaxies 
with stellar mass and redshift from the equation (9) of \citet{schr15} to resolve star formation modes.
They present an analysis of statistical properties of star-forming galaxies
  using the $\it{Herschel}$ and $\it{Hubble}$ $\it{H}$-band images 
  in the redshift range of $\it{z}$ $>$ 0.3.
We extrapolate their relation to 0 $<$ $\it{z}$ $<$ 0.2 bin, but
  found that the extrapolated SFRs are higher than those of previous studies \citep{brin04,elbaz07}. 
Therefore, we used the relation derived from the galaxies at low redshifts
  (i.e. SDSS galaxies z $\approx$ 0.1, \citealt{elbaz07}) to adjust the extrapolated relation;
  we set the average SFR of main sequence to be equal to that of SDSS at log(M$_*$/M$_{\odot}$) = 10.0
  in the redshift range with 0.0 $< \it{z} <$ 0.2, as shown in the left top panel of Figure \ref{fig-sfmode}.  
We define the galaxies within 2 and 0.5 times the average SFRs (dashed lines in Figure \ref{fig-sfmode}) 
as main sequence (MS) galaxies.
Galaxies above the upper dashed line are considered as starbursts (SB, SFR $>$ 2$\times$SFR$_{MS}$),
  and galaxies below the lower dashed line as quiescent galaxies (QS, SFR $<$ 0.5$\times$SFR$_{MS}$).
Our samples are distributed in three different star formation modes at 0.0 $<$ $\it{z}$ $<$ 0.2,
 however there are fewer quiescent galaxies as redshift increases because of the MIR detection limit.
 Note that quiescent galaxies becomes much fainter in the MIR ranges at higher redshift.
Therefore, we constrain the galaxies mass range of 9.0 $<$ log(M$_*$/M$_{\odot}$) $<$ 11.5 as total sample 
  to avoid extreme mass range of galaxies and select the uniform sample over the star formation mode. 

\begin{figure*}
\centering
\includegraphics[width=0.6\textwidth]{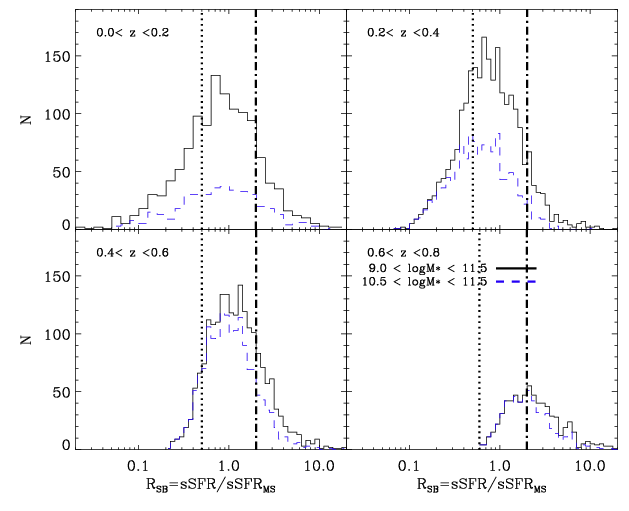}
\caption{Starburstiness R$_{SB}$ distribution of galaxies at each redshift range. Black solid and blue dashed line represent the galaxies in the stellar mass range of 9.0 $<$ log(M$_*$/M$_{\odot}$) $<$ 11.5 and  10.5 $<$ log(M$_*$/M$_{\odot}$) $<$ 11.5, respectively.
Dotted and dash-dotted line represent the borders between starbursts and main sequence, 
main sequence and quiescent galaxies, respectively.}
\label{fig-sbn}
\end{figure*}

To better understand the overall star formation activity for galaxies 
  by minimising the mass effects, we plot the starburstiness (R$_{SB}$) distribution in Figure \ref{fig-sbn}.
Starburstiness represents the star formation activity which is a measure of the excess in specific 
star formation rate of a galaxy compared to that of a main sequence
  galaxy with the same stellar mass and is defined as R$_{SB}$ = sSFR/sSFR$_{MS}$ \citep{elbaz11}. 
Figure \ref{fig-sbn} displays R$_{SB}$ of galaxies in 
  total mass range 9.0 $<$ log(M$_*$/M$_{\odot}$) $<$11.5 (black solid histogram)
  and those at 10.5 $<$ log(M$_*$/M$_{\odot}$) $<$11.5 (blue dashed histogram).
The galaxies with R$_{SB}$ $<$ 0.5 and 2 $<$ R$_{SB}$ represent quiescent and starburst systems, respectively.
As expected, both samples show peaks around R$_{SB}$ =1.
However, the bin of 0.6 $<$ $\it{z}$ $<$ 0.8 has fewer quiescent and 
  main sequence galaxies than other bins because of detection limit.
Therefore we remove the sample in 0.6 $<$ $\it{z}$ $<$ 0.8 for further analysis.
Because the quiescent galaxies could be still affected by detection limits 
  at all redshifts bins except 0.0 $<$ $\it{z}$ $<$ 0.2,
  it should be noted that the merger fractions of quiescent galaxies mean upper limits.

\begin{figure*}
\centering
\includegraphics[width=0.7\textwidth]{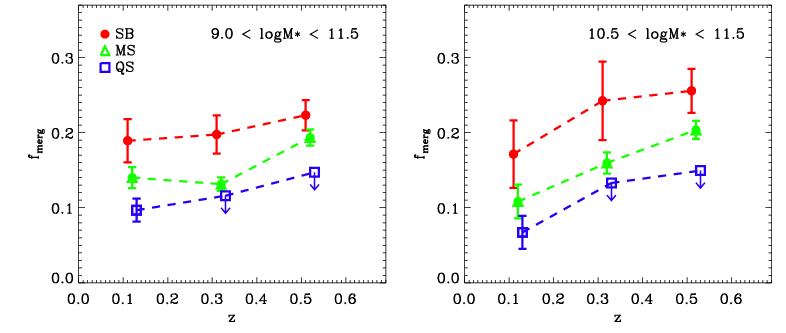}
\caption{Evolution of the merger fractions for starburst, main sequence, and quiescent galaxies.
Filled red circle, open green triangle, open blue rectangle represent starbursts, main sequence, and quiescent galaxies, respectively.
Left panel shows galaxies in the entire stellar mass range, and right panel shows the galaxies in 10.5 $<$ log(M$_*$/M$_{\odot}$) $<$11.5.
}
\label{fig-merg}
\end{figure*}

Figure \ref{fig-merg} shows the evolution of merger fractions for starbursts, 
  main sequence, and quiescent galaxies as a function of the redshift.
We define the merger fraction as the ratio of a number of merging galaxies
  to total number of galaxies in each star formation mode within the redshift range.
To minimise the mass effects on the comparisons of merger fractions between the samples,
  we examine the trend of galaxies with total (9.0 $<$ log(M$_*$/M$_{\odot}$) $<$ 11.5)
  and narrow (10.5 $<$ log(M$_*$/M$_{\odot}$) $<$ 11.5) mass range 
  in the left and right, respectively.
We find that merger fractions of all three different modes of 
  galaxies marginally increase with redshift
  in both panels of different mass ranges.
The merger fractions of galaxies in the total mass range at 0.0 $<$ $\it{z}$ $<$ 0.2 
  are higher compared to those of galaxies in the narrow mass range.
This is because there are more galaxy samples in the log(M$_*$/M$_{\odot}$) $<$ 10.5
  as shown in Fig \ref{fig-sfmode}.
We also find that the merger fractions of galaxies differ for three star formation modes, 
  and the merger fractions of starbursts are higher than those of
  main sequence and quiescent galaxies in both panels.

We fit the merger fractions evolution with power-law \citep{patt02,cons09},
 which is given by f$_{m} = \alpha (1+z)^m + \rm{C}$. 
We use six points of merger fraction with binsize 0.1 of redshift. 
For starbursts and main sequence galaxies, 
  we obtain the index $\it{m}$ = 0.90 $\pm$ 0.18 and 2.04 $\pm$ 0.13 in total mass range, respectively,
  and $\it{m}$ = 1.81 $\pm$ 0.12 and 1.21 $\pm$ 0.08 in narrow mass range, respectively.
These are relatively similar or lower than those of others \citep{cons03,lotz08,qu17}.

To examine whether our results are robust against different main sequence selections,
  we also use the evolutionary trend of main sequence locus in \citet{pear18}.
Following the single power-law they used,
  S = $\alpha$ [log(M$_*$) + 10.5] + $\beta$ \citep{pear18,whit12},
  where $\alpha$ and $\beta$ is the slop and the normalisation, respectively,
  we calculate the fit of SFR and $M_*$ of galaxies.
We fix the $\alpha$ as 0.5 and interpolate the $\beta$ using the parameters 
  from the table 2 in \citet{pear18},
  and identify starbursts, main sequence and quiescent galaxies.
We analyse merger fractions of galaxies and find that
  the increase trends of merger fractions for galaxies
  in different star formation modes as the redshift increase,
  when we use both the average SFRs of \citet{schr15} and \citet{pear18},
  are consistent.
  

\subsection{Merger fraction of galaxies with and without $\it{Herschel}$ detections}
It is well known that FIR bright galaxies tend to be found as mergers at low redshifts. However, this is not always true for high redshift galaxies; isolated disk galaxies at high redshifts can have high infrared luminosities without any merger events because of their large amount of gas \citep{drew20}. This suggests that the infrared luminosity may not reflect genuine physical conditions of galaxies when comparing galaxies at different redshifts. Instead, it is important to distinguish galaxies based on more physically motivated parameters including star formation mode, which is the main driver of this study. To better justify this point, we further examine the merger fraction of FIR detected galaxies depending on the redshift and star formation modes with that of FIR non-detected galaxies.
Here, the FIR detection means that the galaxies are 
  detected at least one band of $\it{Herschel}$/SPIRE 250, 
  350 and 500 $\micron$ wavelengths. 
Because the $\it{Herschel}$/PACS covers only the 
 NEP-Deep field unlike the $\it{Herschel}$/SPIRE (see Fig. 1 in \citealt{kim21}),
 we use the only the $\it{Herschel}$/SPIRE data to reduce the selection effect for the comparison.
  
\begin{figure*}
\centering
\includegraphics[width=0.85\textwidth]{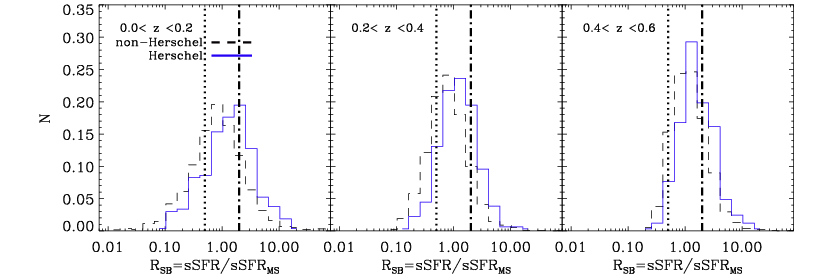}
\caption{Distribution of starburstiness for galaxies at each redshift range. 
Black dashed line represents non-$\it{Herschel}$ detected galaxies 
  and blue solid line represents $\it{Herschel}$ detected galaxies.
}
\label{fig-herschelsbn}
\end{figure*}

\begin{figure*}
\centering
\includegraphics[width=0.70\textwidth]{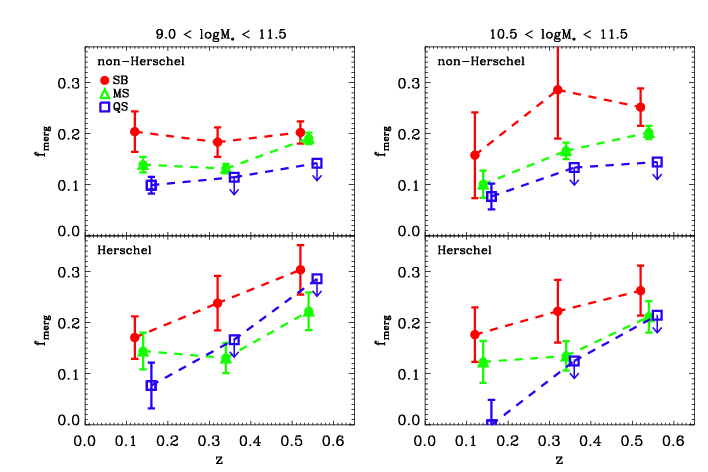}
\caption{Evolution of the merger fraction of starburst, main sequence, and quiescent galaxies
for non-$\it{Herschel}$ detected galaxies (upper) and $\it{Herschel}$ detected galaxies (bottom).
The left and right panels show galaxies with total and fixed mass range of 
   9.0 $<$ log(M$_*$/M$_{\odot}$) $<$ 11.5 and 10.5 $<$ log(M$_*$/M$_{\odot}$) $<$ 11.5.
Coloured symbols are same as Fig \ref{fig-merg}. }
\label{fig-herschel}
\end{figure*}

We separate our 9 $\micron$ detected samples into $\it{Herschel}$ non-detected and detected ones, 
  and show their starburstiness at different redshift bins in Figure \ref{fig-herschelsbn}.
Black dashed and blue solid lines represent
  $\it{Herschel}$ detected and non-$\it{Herschel}$ detected sample, respectively.
We find that $\it{Herschel}$ detected samples have higher R$_{SB}$
 than those of non-$\it{Herschel}$ detected sample in all redshift bins as we can expect.
Figure \ref{fig-herschel} shows the evolution of galaxy merger fraction
  for non-$\it{Herschel}$ detected and $\it{Herschel}$ detected samples 
   in upper and bottom panels, respectively.
Right and left panels show the total and narrow mass range, respectively.
In the total mass range, we find that the merger fraction of starburst galaxies with
  $\it{Herschel}$ detections seemingly increase as the redshift increases
  compared to those of non-$\it{Herschel}$ detected galaxies.
Also, the merger fraction of those with $\it{Herschel}$ detections is higher 
  than those of non-$\it{Herschel}$ detected galaxies,
  because of $\it{Herschel}$ detected galaxies have higher FIR luminosities.
  
We fit the merger fractions evolution with power-law \citep{patt02,cons09},
 which is given by f$_{m} = \alpha (1+z)^m + \rm{C}$.
For non-$\it{Herschel}$ detected galaxies, 
  we obtain the index $\it{m}$ = 0.18 $\pm$ 0.62 and 0.81 $\pm$ 0.19 in total mass range
  and $\it{m}$ = 0.46 $\pm$ 2.29 and 1.44 $\pm$ 2.12 in narrow mass range
  for starbursts and main sequence galaxies, respectively.
For $\it{Herschel}$ detected galaxies, 
  we obtain the index $\it{m}$ = 2.22 $\pm$ 0.72 and 1.09 $\pm$ 0.78 in total mass range
  and $\it{m}$ = 2.71 $\pm$ 2.46 and 1.78 $\pm$ 2.04 in narrow mass range
  for starbursts and main sequence galaxies, respectively.
The indices for merger fractions of 
  starbursts in $\it{Herschel}$ detected galaxies
  are significantly different compared to those in non-$\it{Herschel}$ detected galaxies.
The comparison of them are such as 
  $\it{m}$ = 0.18 $\pm$ 0.62 vs. $\it{m}$ = 2.22 $\pm$ 0.72 in total mass range,
  and $\it{m}$ = 0.46 $\pm$ 2.29 vs. $\it{m}$ = 2.71 $\pm$ 2.46 in narrow mass range.
The differences of main sequence galaxies are relatively weak compared to those of starbursts
  such as $\it{m}$ = 0.81 $\pm$ 0.19  vs. $\it{m}$ = 1.09 $\pm$ 0.78 in total mass range,
  and $\it{m}$ = 1.44 $\pm$ 2.12 vs. $\it{m}$ = 1.78 $\pm$ 2.04 in narrow mass range.

We also examine that the $\it{Herschel}$ detected samples have large range of SFRs at low redshift,
  and the SFRs of those galaxies become higher as the redshift increases.
Thus, the increase tendency of merger fractions for $\it{Herschel}$ detected samples
  may includes the cosmic evolution.
To better compare the merger fractions between $\it{Herschel}$ detected and non-$\it{Herschel}$ detected galaxies 
  by minimising the mass effects,
  we compare the results in narrow mass range in the right panels of Figure \ref{fig-herschel}.
Although errors are large,
  we find that the fraction at the same star formation mode is not different 
  depending on the $\it{Herschel}$ detection considering the errors.
Of course, the merger fractions of different star formation modes 
  are still different for both samples.
This comparison shows the importance of star formation mode in determining the merger fraction
  regardless of FIR luminosities.

\section{DISCUSSION}\label{discuss} 
\subsection{Merger fractions over the star formation modes and their evolution}
The evolution of galaxy merger fraction over the cosmic time has been examined through numerical simulations and observational analysis. Some simulations assuming cold dark matter Universe
  predicted a decreasing merger fraction of galaxies with cosmic time \citep{fakh08,rodr15},
  while other simulations show that the increasing of merger fraction to $z$ $\sim$ 1.5 and then constant
  as redshift increases \citep{kavi15,qu17,snyd17}.
In observations, \citet{cons08} showed that the merger fraction of very massive galaxies 
  with log(M$_*$/M$_{\odot})>$ 10 appears to increase up to $\it{z}$ $\sim$ 3,
  while the merger fraction of less massive galaxies has a peak  $\it{z}$ $\sim$ 1.5 -- 2.5 
  and decreases to high redshift.  \citet{vent17} also showed that the merger fractions for galaxies 
  with log(M$_*$/M$_{\odot}) >$ 9.5 increase to around $\it{z}$ $\sim$ 2 and slowly decrease after that.
Even in relatively low redshift range, some authors found an increasing merger fraction with redshift \citep{lope09,vent17}, however constant merger fraction at $\it{z}$ $<$ 0.6 is also suggested \citep{cons09,joge09}.
Our results are in agreement with the observations suggesting that
 the merger fraction of galaxies slightly increases with redshift $\it{z}$ $<$ 0.6 \citep{lope09,man16,vent17}.
However, the absolute values of merger fractions could differ due to different methods of sample selections depending on luminosity, mass or definition of a merger, which will be discussed in Section \ref{method}.

As shown Figure \ref{fig-merg}, merger fractions of starburst, main sequence and quiescent galaxies are dependent of star formation mode. Although some authors suggested the dependency of merger fraction of galaxies on the distance from the main sequence \citep{cibi19,pear19}, the effect of star formation modes could not be evaluated quantitatively.
For the fair comparison, we try to investigate the evolution of merger fraction for galaxies with similar star formation activities. 

Merger galaxies selected by morphology are mainly late-stage and disturbed systems  \citep{pear19}.
Because our merger galaxy samples are also selected by morphology, 
  the higher merger fraction in this study than those in other studies can suggest that
  the star formation enhancement is prominent at the late stage of merging \citep{sand96,haan11,cox06,hwang12}
  and earlier stage of merging only cause mild increase of SFRs for close pairs \citep{lin07}.

Regarding the evolution of merger fraction, \citet{cons09} showed diverse results through the fitting with a power-law function. They found that the power-law slope changes from 1.5 to 3.8  depending on sample selection and different merger fraction at $\it{z}$ = 0.
The slope tend to be higher for more massive galaxies and lower for less massive galaxies \citep{cons03}.
\cite{qu17} used an exponential power-law function for the simulation predictions,
  they found that the power-law slope, $\it{m}$, for close pairs with $\it{z}$ $<$ 4
  changes from 2.8 to 3.7 depending on the mass limits.
Otherwise, others who used morphological disturbances for merger selections 
 and the redshift range of galaxies with $\it{z}$ $<$ 1.2 \citep{lotz08}
 obtained the mild slope of $\it{m}$ = 1.26.
Considering similar merger selection and redshift range, our results on the power-law slope of $\it{m}$ = 0.18 -- 2.71
 are consistent with those from \cite{lotz08}.
However, they showed that the slope of merger fraction could be easily 
  affected by morphological diagnostics and  timescales to determine merger fractions.

\subsection{Mergers for $\it{Herschel}$ detected galaxies}
Galaxy merging is expected to drive star formation episodes \citep{barn96,miho96},
  however, UV/optical light is dimmed and sources appear redder due to absorption and scattering by dust.
Since considerable amounts of the energy from star formations and AGNs have been absorbed by gas and dust and re-emitted in FIR wavelengths \citep{puge96,dole06}, FIR data set would be good for the study of star formation activity \citep{pei99,chary01}.
Some results for LIRGs and ULIRGs showed that FIR-bright galaxies are ongoing mergers and have disturbed morphology, which are the evidence for merger activities \citep{sand88,clem96,hopk06,hwang10}.
While these studies were mainly focused on FIR-bright galaxies, our samples selected from MIR detections have a wider range of L$_{TIR}$. As shown in Figure \ref{fig-merg} and \ref{fig-herschel}, the merger fractions are strongly dependent of star formation modes, irrespective of $\it{Herschel}$ detection. 
We also find that the increasing slope of merger fraction for starbusrts detected in $\it{Herschel}$ is steeper  
  than that of non-$\it{Herschel}$ detected starbursts.
The difference of the slope for main sequence galaxies is not significant compared to those of starbursts.
Note that quiescent galaxies shows the steepest slope, however merger fractions at $\it{z}$ $>$ 0.2 are upper limit due to the lack of sample.
These results could support that $\it{Herschel}$ detected galaxies with high FIR luminosity such as LIRGs/UIRGs are more stochastically in the merging stage. 
Although it is difficult to compare with other results in the effectiveness of FIR detection,
 this can be interpreted that the merger fractions of galaxies are determined not only by the IR luminosity, but also by the star formation mode of galaxies at fixed redshift range.
  
\subsection{Comparison to other studies}\label{method}
To study the merger fraction of galaxies, 
  one has to define a galaxy sample along with redshift/mass range 
  and galaxy classification method \citep{lotz08,cons09,bund09,man16,dela17,wats19}.
Therefore, it is important to understand the sample selection
  including the merger identification scheme to make fair comparison with other studies.
Although it is difficult to directly compare our results with other studies
  because of these differences, we describe the similarity and 
  the difference between our study and other studies in this section.

Methodologically, merger galaxies can be identified by
  using galaxy pairs or morphological disturbances.
As morphological cases,
\citet{lotz08} used Gini and M$_{20}$ for selecting merger galaxies 
  in 0.2 $<$ $\it{z}$ $<$ 1.2 and volume limited sample with 
  B-band luminosity limits assuming the luminosity evolution.
They found weak evolution of the merger fractions of galaxies
  in this redshift range.
\citet{cons09} derived the increasing merger fraction using asymmetry and clumpiness 
  with galaxies from COSMOS and EGS between 0.2 $<$ $\it{z}$ $<$ 1.2.
Although there are differences between our sample and theirs
  such as redshift range and existence of MIR data, 
  our result is consistent with the previous ones \citep{lotz08,cons09}
  that the merger fractions for galaxies mildly increase as redshift increases
  with using morphological selection for merger galaxies.
  
In addition to morphological method, there are studies of merger fractions using galaxy pairs.
 \citet{bund09} and \citet{dera09}  use mass-selected pairs and
  projected separation (R$_{proj}$) for selecting merger galaxies, respectively. 
\citet{wats19} showed that
  the merger fraction for paired galaxies in clusters is higher
  than those in field environments \citep{bund09,dera09}.
\citet{cibi19} compared the results from the morphological classification
  with those from the pair identification.
They found that most of the starbursts galaxies are morphologically disturbed,
   but for galaxy pairs, the merger fractions were small in starburst galaxies.
Thus, this can suggest that 
  the merger fractions of this study could be higher than 
  that in other studies based on galaxy pairs.

Relatively high merger fractions of our results also
 can be explained by sample selection criteria.
\citet{lotz08} used luminosity-size limits for selecting of massive galaxies.
\citet{cons09,lope09} used galaxies with M$_{*} >$ $10^{10}$ M$_{\odot}$.
These criteria secure limited galaxies compared to our sample that
  have mass range of 9.0 $<$ log(M$_*$/M$_{\odot}$) $<$11.5.
However, the largest difference of the sample selection between ours and others is
   the use of the MIR detection in our study,
   which can significantly affect the star formation activity. 
Then, the number of sample can be limited, 
  this small total number of sample which is denominators of merger fractions
  could affect that the merger fractions of galaxies become high compared to others.
The method also can affect results;
  because Gini-M$_{20}$ are sensitive to the features of minor mergers \citep{lotz11},
  our method may select more merger candidates
  than other studies based on CAS or asymmetry criteria.
\citet{pear19} reported elevated merger fraction for galaxies at 0 $< \it{z} <$ 4
  based on CANDELS data compared to other studies.
Such a result may be arisen
  because the pixel size of galaxies within the images 
  becomes smaller and galaxies become fainter 
  as redshift increases,
  then the suppressed galaxy features counted as merger galaxies. 
Therefore, the direct comparison of absolute values of merger fractions
  between different studies is difficult.

 \section{SUMMARY}\label{sum}
 We used the galaxy sample detected at the MIR band (9 \micron) of AKARI 
  in the NEP--Wide field. In order to identify the merging galaxies, the morphological analyses were carried out 
  relying on the Gini and M$_{20}$ coefficients derived from deep Subaru/HSC ($\it{r}$-, $\it{i}$-band) images.
Using the spectroscopic and photometric redshifts, we derived total infrared luminosity and SFR from AKARI 9 $\micron$ detections. We compare the merger fractions between three different star formation modes at $\it{z} <$ 0.6:
  starburst, main sequence and quiescent galaxies.
Our main results are as follows:

\begin{enumerate}
\item
The merger fractions for starbursts, main sequence, and quiescent galaxies slightly increase with redshift at $\it{z} <$ 0.6.

\item 
The galaxy merger fractions differ depending on the star formation mode.
The starbursts show higher merger fractions than those of main sequence and quiescent galaxies.

\item 
The increasing slope of the merger fractions for $\it{Herschel}$ detected starbursts slightly steeper compared to those of non-$\it{Herschel}$ detected starbursts.
 
\end{enumerate}
 
Our results are in line with the idea that the merger fraction increases with redshift in local Universe \citep{lotz08,cons09,lope09} and galaxies in the different star formation modes such as starbursts,
  main sequence and quiescent galaxies show different merger fractions \citep{cibi19,pear19}.
Regardless of the FIR detection, the increasing trends of the merger fraction over local universe ensure the consistency in all the different star formation modes.
These results underscore the importance of the star formation mode in the study of evolution of galaxy merger fraction.
To better understand the merger fraction evolution with different star formation activities,
  it is important to secure a larger, unbiased sample of high-$\it{z}$ galaxies,
  which does not suffer from observational selection effects on the star formation mode.
 
\section*{Acknowledgements}
W-SJ, EK and Y-SJ acknowledges support from the National Research Foundation of Korea (NRF) grant funded by the Ministry of Science and ICT (MSIT) of Korea (NRF-2018M1A3A3A02065645). HSH was supported by the New Faculty Startup Fund from Seoul National University.
HShim acknowledges the support from the National Research Foundation of Korea grant No. 2018R1C1B6008498.
TH is supported by the Centre for Informatics and Computation in Astronomy (CICA) 
at National Tsing Hua University (NTHU) through a grant from the Ministry of Education 
of the Republic of China (Taiwan).

\section*{Data Availability}

The data underlying this article will be shared on reasonable request to the corresponding author.



\end{document}